\documentclass{iau} 
\usepackage{graphicx}

\newcommand{\HI}{{H}{\sc i}}
\newcommand{\sunn}{$_{\odot}$}

\newcounter{qub}

\title[Nearby void dwarf galaxies] 
{Nearby void dwarf galaxies: recent results, \\
the ongoing project and prospects}

\author[S.A.~Pustilnik, D.I.~Makarov \& A.L.~Tepliakova]
{Simon A.~Pustilnik$^1$, Dmitri I.~Makarov$^1$, \and Arina L. Tepliakova$^1$}

\affiliation{$^1$Special Astrophysical Observatory of RAS, \\ 369167,
Nizhnij Arkhyz, Karachai-Circessia, Russia \\
email: {\tt sap@sao.ru, dim@sao.ru}}

\pubyear{2019}
\volume{344}
\jname{Dwarf Galaxies: From the Deep Universe to the Present}
\editors{K.~McQuinn, S.~Stierwalt, eds.}

\begin{document}

\maketitle

\begin{abstract}
Properties of dwarf galaxies formed and evolved in the lowest density
environment remain largely unexplored and poorly understood. Especially
this concerns the low-mass end ($M_{\rm bar} < 10^9$ M\sunn). We overview
the results of systematic study of a hundred void dwarfs from the nearby
Lynx-Cancer void. We describe the ongoing project aiming to form Nearby Void
galaxy sample ($R < 25$ Mpc) over the whole sky. 1354 objects with distances
less than 25 Mpc fall within 25 voids delineated by 460 luminous
galaxies/groups. The void major sizes range from 13 to 37 Mpc.
1088 of 1354 void galaxies reside deeply in voids, having distances to
the nearest luminous neighbour of 2--11 Mpc. 195 nearest void galaxies
reside in the Local Volume. We summarize the main statistical properties
of the new sample and outline the prospects of study of
both, the void dwarf properties and the fine structure of voids.
\end{abstract}

\firstsection 
\section{Introduction}

Study of voids and their galaxies became especially popular in XXI century
thanks to the massive redshift data from several very large surveys, such
as SDSS, 2dFGRS, 6dF, ALFALFA,  and the large progress in models and
simulations.

However, almost all void galaxy studies are biased to distant (80--200 Mpc)
voids, and hence, to the upper part of the void galaxy Luminosity Function
($M_{\rm B}$, $M_{\rm r} < -16$).
Some models hint on possible delay in the epoch of formation and on the slow
evolution of small void galaxies (e.g.,
\cite[Einasto et al. (2011)]{Einasto2011},
\cite[Aragon-Calvo \& Szalay (2013)]{AragonCalvoSzalay2013}).
However, observationally void galaxies with
$M_{\rm B} > -14^m$, and M$_{\rm bar} \lesssim 10^9$ M\sunn\ remain almost
unexplored.

The main goal of our ongoing project is to fill the almost empty niche of
'Galaxies in the Nearby Voids', with the emphasis on studies of the smallest
void galaxies. The first step in this direction is the formation of sample of
nearby void galaxies. See the paper submitted to MNRAS (Pustilnik, Tepliakova,
Makarov "Void galaxies in the nearby Universe. I. Sample description").

\section{Overview of results on galaxies in the nearby Lynx-Cancer void}

The Lynx-Cancer void was first described in detail by
\cite[Pustilnik \& Tepliakova (2011)]{PistilnikTepliakova11}. Its
center is situated at the distance of 18 Mpc. The void's volume is $\sim$2000
cub.~Mpc, it contains 108 galaxies with $M_{\rm B}$ of --9.6$^m$ to
--18.4$^m$ and with M(\HI) of $10^6$ to $5\times10^9$ M\sunn. The galaxy
sample is complete to $M_{\rm B} \sim -14.0^m$, almost all
void galaxies are late-type, and about a half of them are low surface
brightness dwarfs (LSBDs).

The main conclusions from the study of relation "O/H versus $M_{\rm B}$"
(\cite[Pustilnik, Perepelitsuna \& Kniazev (2016)]{Pustilniketal16}, see also
\cite[Kniazev, Egorova \& Pustilnik (2018)]{Kniazevetal18} )
and the ratio of M(\HI)/$L_{\rm B}$ (\cite[Pustilnik \& Martin (2016)]
{PustilnikMartin16}), and the comparison with the control samples in
the Local Volume are as follows.

For void galaxies we see that: \\
a) The gas O/H in average is lower by $\sim$40\% (on 81 galaxies), \\
b) M(\HI) in average is larger by $\sim$40\% for a fixed $M_{\rm B}$
   (on 103 galaxies), \\
c) Several fainter/faintest void galaxies appear the extreme outlyers in
   O/H (by 2--5 times lower than for the control sample) and/or the gas
   mass-fraction (up to 98--99 per cent)
(e.g., \cite[Pustilnik, et al. (2010)]{Pustilniketal10},
\cite[Chengalur \& Pustilnik (2013)]{Chengalur13},
\cite[Chengalur, Pustilnik \& Egorova(2017)]{Chengalur17}),
\\
d) Colours of the galaxies' outer parts versus the PEGASE evolutionary tracks
   are used as the age indicators. The great majority of galaxies have visible
 population with $T_{\rm stars} \sim 10-14$ Gyr, and hence, have been evolving
 during the cosmological time. However, for $\sim$15\%, the colours correspond
 to $T_{\rm stars} \sim 1-5$ Gyr
(\cite[Perepelitsyna, Pustilnik \& Kniazev (2014)]{Perepeletal14}). \\
These findings: 1)  imply the slower evolution of void galaxies as a whole,
and 2) hint on the presence of void galaxies with much delayed formation
epoch, and in particular, on the presence of half-dozen void VYGs (Very Young
Galaxies, as defined by \cite[Tweed et al. (2018)]{Tweed18}, following
the discovery of the record-low metallicity star-forming dwarf J0811+4730
by \cite[Izotov et al. (2018)]{Izotovetal18}).

\section{Main parameters of the Nearby Voids and their galaxies}

The main goal of the ongoing project is to form a large, deep and
representative sample of dwarf galaxies residing in voids of the nearby
Universe. The sample is the basement for the mass study of the galaxy content,
their evolutionary status, clustering and dynamics with respect to their
counterparts residing in more typical, denser regions and for study of
void small-scale substructures.

\begin{figure}[b]
\begin{center}
 \includegraphics[width=3.0cm,angle=-90,clip=]{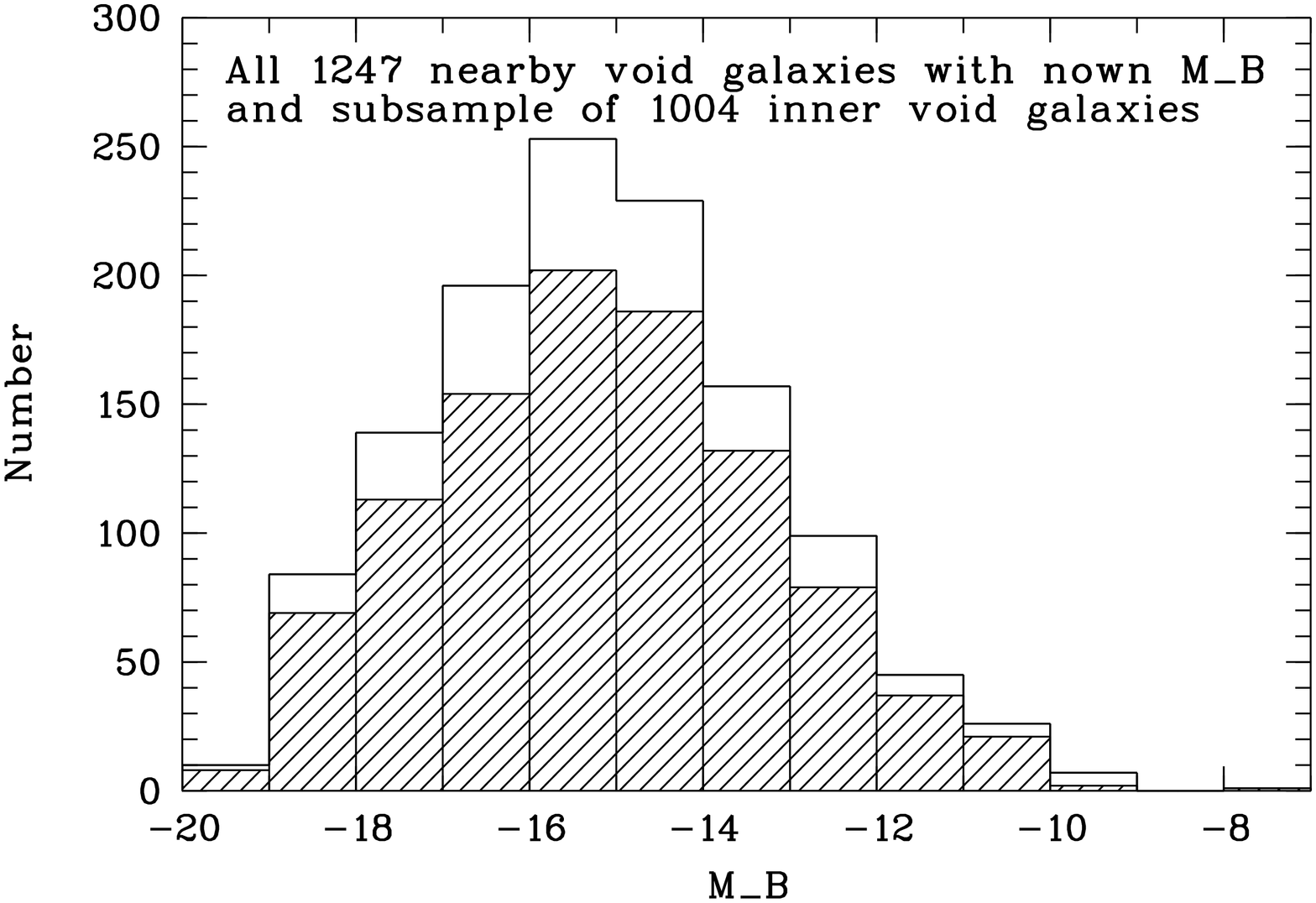}
 \includegraphics[width=3.0cm,angle=-90,clip=]{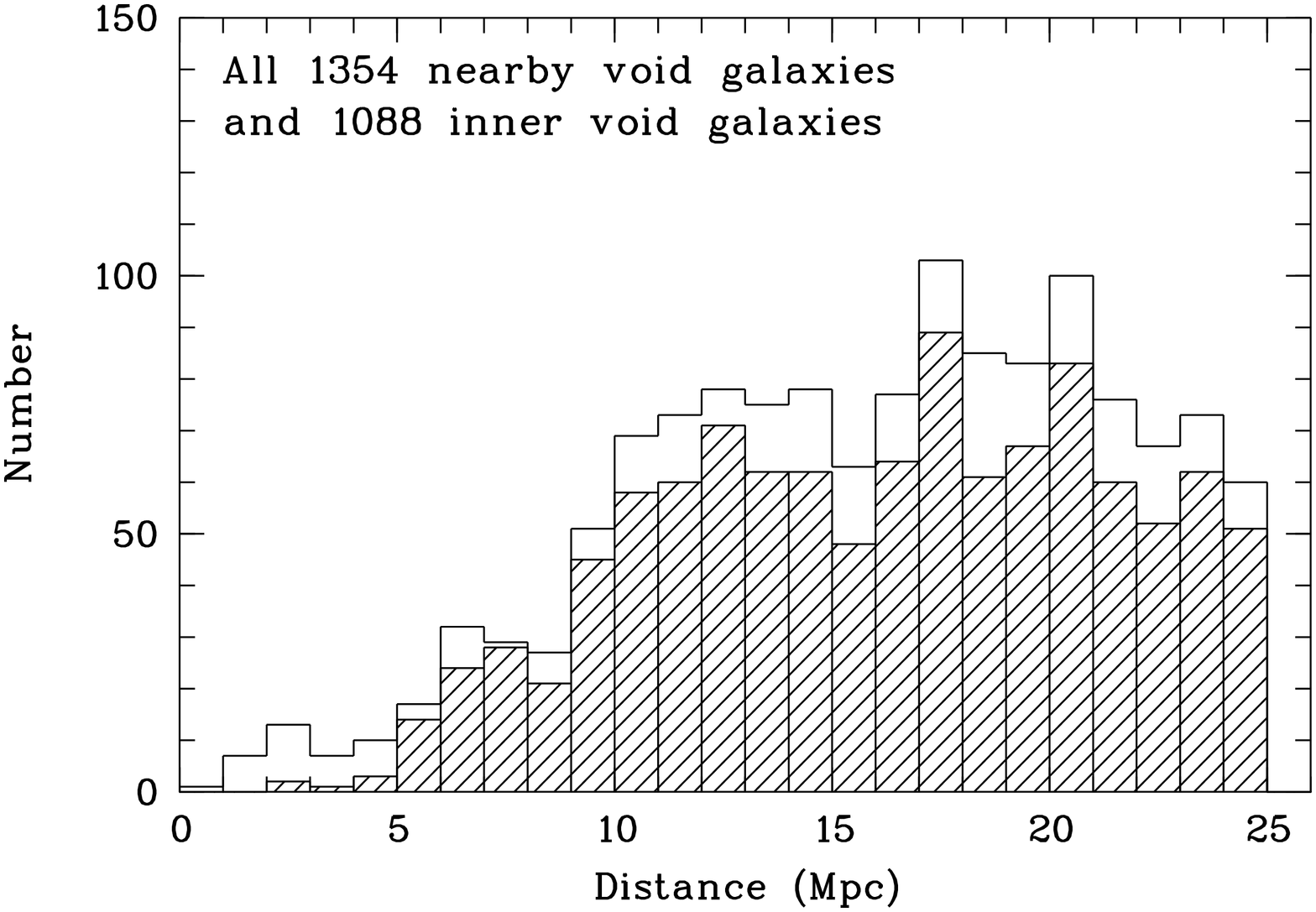}
 \includegraphics[width=3.0cm,angle=-90,clip=]{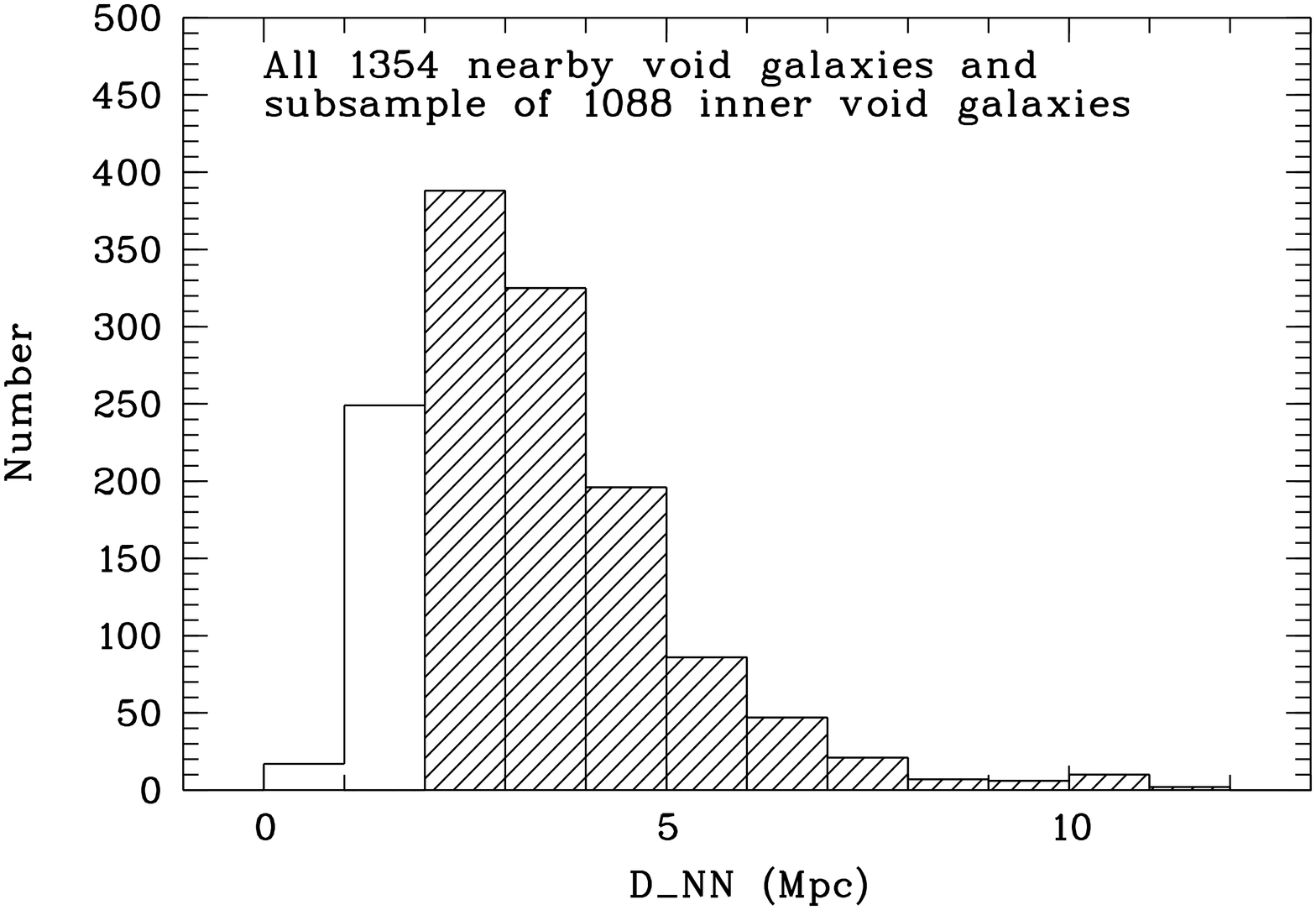}
\caption{Distributions of all void galaxies and 'inner' (hatched) void sample.
   {\bf Left panel}. $M_{\rm B}$ distribution of all 1247 void galaxies with
known $M_{\rm B}$, with median of -15.1$^m$. {\bf Middle panel}. Distance
 distribution of all 1354 void galaxies. {\bf Right panel}. Distribution of
distances of void galaxies to the nearest luminous galaxy/group $D_{\rm NN}$.}
   \label{fig1}
\end{center}
\end{figure}

We identify 25 voids over the entire sky within the distance of $R = 25$ Mpc
from the Local Group. The voids are defined as groups of lumped empty spheres
bounded by
$\sim$450 `luminous' galaxies (or groups with `luminous' galaxies) with the
absolute $K$-band magnitudes $M_{\rm K} < -22.0$.
1354 fainter galaxies residing in these nearby voids have absolute blue
magnitudes of $M_{\rm B}$ in the range of --7.5$^m$ to --19.5$^m$, with the
median of --15.1$^m$. Of them, 1088 belong to the  'inner' subsample, with the
distances to the nearest luminous neighbour of $D_{\rm NN} \geq$ 2.0 Mpc.
Of them, 195 galaxies fall within the Local Volume ($R < 11$ Mpc).
The distributions of the Nearby Void galaxies on $M_{\rm B}$, distances from
the Local Group and $D_{\rm NN}$ are shown Fig.~\ref{fig1}, where the hatched
part of the histograms separates the 'inner' subsample.

In Table~\ref{tab1} we present the list of the Nearby Voids along with their
main parameters. It includes the void number in the catalog, the approximate
equatorial
coordinates and the distance (in Mpc) of void center, the maximal extent of
void in three projections in supergalactic coordinats, the total number of
galaxies assigned to this void and the respective number of galaxies from
the 'inner' void sample, that is galaxies with $D_{\rm NN} \geq$ 2.0 Mpc.

\begin{table}
\begin{center}
\caption{Main parameters of nearby voids}
\label{tab1}
\vspace{0.1cm}
{\bf {\scriptsize
\begin{tabular}{|r|l|l|l|c|c|c|c|} \hline  \\[-0.2cm]
{\#}     & {\bf Void name}   & {\bf RA$_{\rm c}$} & {\bf Dec$_{\rm c}$}  &
{\bf Dist$_{\rm c}$}  & {\bf Max.ext.}  & {\bf Tot.}  & {\bf Inner}  \\
& {  }  & {hours}  & {degr}  & {Mpc }  &{$\Delta X$,$\Delta Y$,$\Delta Z$}  &
{void}  & {void}  \\
&{  }  &{  }  & {  }  &{  }  & {Mpc}  & {gals}  & {gals}  \\
& { (1) }  &{ (2) }  &{ (3) }  &{ (4) }  &{ (5) }  & { (6) }  &{ (7) }  \\
\\[-0.2cm] \hline \\[-0.2cm]
  1  & Cas-And     &   00.7 & +53  &    19.0 &19,19,23  & 19  &  15    \\
  2  & Tuc         &   00.9 & --64 &    11.2 &14,14,14  & 56  &  44    \\
  3  & Cet-Scu-Psc &   01.3 & --02 &    15.2 &33,17,29  &108  &  85    \\
  4  & Pho         &   01.4 & --54 &    18.0 &17,19,18  & 80  &  66    \\
  5  & Tau         &   03.8 & +17  &    18.8 &24,27,21  & 53  &  46    \\
  6  & Per         &   04.0 & +52  &    19.7 &14,14,13  &  4  &   3    \\
  7  & Eri-Ori     &   05.1 & --07 &    18.5 &20,18,17  & 41  &  30    \\
  8  & Ori-Tau     &   05.4 & +15.2&    07.5 &18,14,14  & 46  &  36    \\
  9  & Aur         &   05.8 & +38  &    13.5 &23,22,21  & 36  &  34    \\
 10  & Lep         &   05.85 &--17 &    06.3 &13,13,13  & 43  &  32    \\
 11  & Mon         &   06.4 &--07  &    20.2 &19,15,16  &  7  &   2    \\
 12  & Cnr-CMi-Hyd &   08.5 & +10  &    17.5 &29,25,23  & 129 & 106    \\
 13  & Vel         &   09.5 &--50  &    19.0 &20,27,21  & 76  &  59    \\
 14  & Hyd         &   09.8 &--15.2&    19.2 &18,14,17  & 22  &  15    \\
 15  & Cen-Cir     &   14.4 &--65  &    21.0 &15,14,15  & 20  &  14    \\
 16  & UMa         &   14.8 &+59   &    21.0 &15,16,13  & 82  &  73    \\
 17  & Vir-Boo     &   14.8 &+07   &    10.2 &14,14,15  & 20  &  12    \\
 18  & Boo         &   15.3 & +27  &    19.4 &20,24,26  & 49  &  33    \\
 19  & Lib         &   15.4 &--26.5&    18.3 &19,21,24  & 40  &  31    \\
 20  & Her         &   16.6 &+20   &    13.5 &21,22,25  & 118 &  97    \\
 21  & Oph-Sgr-Cap &   18.5 &--18  &    13.0 &37,30,35  & 121 &  89    \\
 22  & Dra-Cep     &   20.4 &+71.1 &    13.9 &21,21,17  & 50  &  44    \\
 23  & Cyg         &   20.6 &+36.3 &    19.3 &22,23,28  & 20  &  18    \\
 24  & Pav-Oct     &   20.7 &--73.1&    15.7 &23,20,18  & 36  &  30    \\
 25  & Aqu         &   22.7 &--02.5&    15.5 &20,22,24  & 80  &  72    \\
\\[-0.25cm]
\hline 
\end{tabular}
} }
\end{center}
\end{table}

\section{Summary and prospects}

The absolute majority of the Nearby Void dwarfs are irregular galaxies and
late-type spirals. About 30 blue gas-rich dwarfs are candidates to VYGs.
About 7\%\ of the void sample are of early types: dE/E-S0 galaxies with the
wide range of $M_{\rm B}$. Most of them are well isolated, and thus can
represent an unusual sub-type of field/void early-type galaxies.

The new large sample of 1354 galaxies residing in the Nearby Voids opens
the prospects of systematical studies of galaxy formation and evolution as
well as of the properties of voids themselves. These include
the following directions:

1) search for and study of the lowest mass void dwarfs, including candidates
to the so-called Very Young Galaxies;

2) study the origin and evolution of early-type galaxies in voids;

3) search for appearance of cold accretion as a driving mechanism of galaxy
evolution;

4) study of void small-scale structure, in particular as a probe of
   the predicted effect of voids as 'time machine and cosmic microscope'
 (\cite[Aragon-Calvo \& Szalay (2013)]{AragonCalvoSzalay2013}) and of
   the possible role of Warm Dark Matter in its formation;

5) study the void galaxy motions from their void centers as a mean to
  determine void global properties.

The work is supported by grant of Russian Science Fund No. 14-12-00965.

\end{document}